\documentclass[12pt]{article}
\usepackage{amsfonts}
\usepackage[centertags]{amsmath}
\usepackage{amssymb}
\usepackage[verbose,colorlinks=true,naturalnames=true,linkcolor=blue,]{hyperref}

\def\ba{\begin{array}}
\def\ea{\end{array}}
\def\be{\begin{equation}}
\def\ee{\end{equation}}
\def\bea{\begin{eqnarray}}
\def\eea{\end{eqnarray}}
\usepackage[verbose,colorlinks=true,naturalnames=true,linkcolor=blue,]{hyperref}

\newcounter{rown}

\begin{document}

\title{Quantum deformations of $D=4$ Euclidean, Lorentz, Kleinian and quaternionic $\mathfrak{o}^{\star}(4)$ symmetries\\ in unified $\mathfrak{o}(4;\mathbb{C})$ setting -- Addendum %%\\[10pt]
}
\author{A. Borowiec$^{1}$, J. Lukierski$^{1}$ and V.N. Tolstoy$^{1,2}$ \\
%EndAName
\\
$^{1}$Institute for Theoretical Physics, \\
University of Wroc{\l }aw, pl. Maxa Borna 9, \\
50--205 Wroc{\l }aw, Poland\\
\\
$^{2}$Lomonosov Moscow State University,\\
Skobeltsyn Institute of Nuclear Physics, \\
Moscow 119991, Russian Federation}
\date{}
\maketitle

\begin{abstract}
In our previous paper \cite{BoLuTo2016} we obtained a full classification of nonequivalent quasitriangular quantum deformations for the complex $D=4$ Euclidean Lie symmetry $\mathfrak{o}(4;\mathbb{C})$. The result was presented in the form of a list consisting of three three-parameter, one two-parameter and one one-parameter nonisomorphic classical $r$-matrices which provide 'directions' of the nonequivalent quantizations of $\mathfrak{o}(4;\mathbb{C})$. Applying reality conditions to the complex $\mathfrak{o}(4;\mathbb{C})$ $r$-matrices we obtained the nonisomorphic classical $r$-matrices for all possible real forms of $\mathfrak{o}(4;\mathbb{C})$: Euclidean $\mathfrak{o}(4)$, Lorentz $\mathfrak{o}(3,1)$, Kleinian $\mathfrak{o}(2,2)$ and quaternionic $\mathfrak{o}^{\star}(4)$ Lie algebras. In the case of $\mathfrak{o}(4)$ and $\mathfrak{o}(3,1)$ real symmetries these $r$-matrices give the full classifications of the inequivalent quasitriangular quantum deformations, however for $\mathfrak{o}(2,2)$ and $\mathfrak{o}^{\star}(4)$ the classifications are not full. In this paper we complete these classifications by adding three new three-parameter $\mathfrak{o}(2,2)$-real $r$-matrices and one new three-parameter $\mathfrak{o}^{\star}(4)$-real $r$-matrix. All nonisomorphic classical $r$-matrices for all real forms of $\mathfrak{o}(4;\mathbb{C})$ are presented in the explicite form what is convenient for providing the quantizations. We will mention also some applications of our results to the deformations of space-time symmetries and string $\sigma$-models. 
%in the quantum field theory and gravity. 
\end{abstract}

\setcounter{equation}{0}
\section{Introduction}
The search for quantum gravity is linked with studies of noncommutative space-times and quantum deformations of space-time symmetries. The considerations of simple dynamical models in quantized gravitational background indicate that the presence of quantum gravity effects generates noncommutativity of $D=4$ space-time coordinates, and as well the Lie-algebraic space-time symmetries (e.g. Euclidean, Lorentz, Kleinian, quaternionic and their ingomogeneous versions) are modified into respective quantum symmetries, described by noncocommutative Hopf algebras, named quantum deformations \cite{Dr1986}. Therefore, studing all aspects of the quantum deformations in details is an important issue in the search of quantum gravity models.

For classifications, constructions and applications of quantum Hopf deformations of an universal enveloping algebra $U(\mathfrak{g})$ of a Lie algebra $\mathfrak{g}$, Lie bialgebras ($\mathfrak{g},\delta$) play an essential role (see e.g. \cite{Dr1986}--\cite{Ma1995}). Here the {\it cobracket} $\delta$ is a linear skew-symmetric map $\mathfrak{g}\rightarrow\mathfrak{g}\wedge\mathfrak{g}$ with the relations consisted with the Lie bracket in $\mathfrak{g}$:
\begin{eqnarray}
\begin{array}{rcl}\label{in1}
&&\delta([x,y])= [\delta(x),\Delta_{0}(y)]+[\Delta_{0}(x),\delta(y)],
\\[7pt]
&&(\delta\otimes\mathop{\rm id})\delta(x)+{\rm cycle}=0,
\end{array}
\end{eqnarray}
where $\Delta_{0}(\cdot)$ is a trivial (non-deformed) coproduct
\begin{eqnarray}\label{in2}
\Delta_{0}(x)=x\otimes1+1\otimes x, 
\end{eqnarray}
for any $x,y\in\mathfrak{g}$. The first relation in (\ref{in1}) is a condition of the 1-cocycle and the second one is the co-Jacobi identity (see \cite{Dr1986, Ma1995}). The Lie bialgebra ($\mathfrak{g},\delta$) is a correct infinitesimalization of the quantum Hopf deformation  of $U(\mathfrak{g})$ and the operation $\delta$ is an infinitesimal part of difference between a coproduct $\Delta$ and an oposite coproduct $\tilde{\Delta}$ in the Hopf algebra, $\delta(x)=h^{-1}(\Delta-\tilde{\Delta})\mod h$, where $h$ is a deformation parameter. Any two Lie bialgebras ($\mathfrak{g},\delta$) and ($\mathfrak{g},\delta'$) are isomorphic (equivalent) if they are connected by a $\mathfrak{g}$-automorphism $\varphi$ satisfying the condition 
\begin{eqnarray}\label{in3}
\delta(x)=(\varphi\otimes\varphi)\delta'(\varphi^{-1}(x)) 
\end{eqnarray}
for any $x\in\mathfrak{g}$. Of our special interest here are the quasitriangle Lie bialgebras ($\mathfrak{g},\delta_{(r)}$):=($\mathfrak{g},\delta,r$), where the cobracket $\delta_{(r)}$ is given by the classical $r$-matrix $r\in\mathfrak{g}\wedge\mathfrak{g}$ as follows:
\begin{eqnarray}\label{in4}
\delta_{(r)}(x)=[r,\Delta_{0}(x)].
\end{eqnarray}
It is easy to see from (\ref{in2}) and (\ref{in3}) that \textit{two quasitriangular Lie bialgebras ($\mathfrak{g},\delta_{(r)}$) and ($\mathfrak{g}, \delta_{(r')}$) are isomorphic iff the classical $r$-matrices $r$ and $r'$ are isomorphic, i.e. $(\varphi\otimes\varphi)r'=r$}. Therefore for a classification of all nonequivalent quasitriangular Lie bialgebras ($\mathfrak{g},\delta_{(r)}$) of the given Lie algebra $\mathfrak{g}$ we need to find all nonequivalent (nonisomorphic) classical $r$-matrices. Because nonequivalent quasitriangular Lie bialgebras uniquely determine non-equivalent quasitriangular quantum deformations (Hopf algebras) of $U(\mathfrak{g})$ (see \cite{Dr1986, EtKa1996}) therefore the classification of all nonequivalent quasitriangular Hopf algebras is reduced to the classification of all nonequivalent classical $r$-matrices. 

Let $\mathfrak{g}^{*}:=(\mathfrak{g},*)$ be a real form of a semisimple complex Lie algebra $\mathfrak{g}$, where $*$ is an antilinear involutive antiautomorphism of $\mathfrak{g}$, then \textit{the bialgebra $(\mathfrak{g}^{*},\delta_{(r)})$ is real iff the classical $r$-matrix $r$ is $*$-anti-real ($*$-anti-Hermitian)}.\footnote{All bialgebras over the semisimple complex and real Lie algebras are quasitriangular, due to Whitehead lemma (see e.g. \cite{Ja1979}).} Indeed, the condition of $*$-reality for the bialgebra $(\mathfrak{g}^{*},\delta)$ means that
\begin{eqnarray}\label{in5}
\delta(x)^{*\otimes*}=\delta(x^{*}).
\end{eqnarray}
Applying this condition to the relations (\ref{in4}) we abtain that
\begin{eqnarray}\label{in6}
r^{*\otimes*}=-r,
\end{eqnarray}
i.e. the $r$-matrix $r$ is $*$-anti-Hermitian.

In the previous paper \cite{BoLuTo2016} we obtained a full classification of nonequivalent quasitriangular quantum deformations for the complex $D=4$ Euclidean Lie symmetry $\mathfrak{o}(4;\mathbb{C})$. The result was presented in the form of a list consisting of three three-parameter, one two-parameter and one one-parameter nonisomorphic classical $r$-matrices which provide directions of the nonequivalent quantizations of $\mathfrak{o}(4;\mathbb{C})$. Applying reality conditions in \cite{BoLuTo2016} we obtained the nonisomorphic classical $r$-matrices for all possible real forms of $\mathfrak{o}(4;\mathbb{C})$: Euclidean $\mathfrak{o}(4)$, Lorentz $\mathfrak{o}(3,1)$, Kleinian $\mathfrak{o}(2,2)$ and quaternionic $\mathfrak{o}^{\star}(4)$ Lie algebras. In the case of $\mathfrak{o}(4)$ and $\mathfrak{o}(3,1)$ real symmetries these $r$-matrices give the full classifications of the nonequevalent quasitriangular quantum deformations, however for $\mathfrak{o}(2,2)$ and $\mathfrak{o}^{\star}(4)$ the classifications are not complete. In this paper we provide the full classifications of the quantum deformations in the case of Kleinian $\mathfrak{o}(2,2)$ and quaternionic $\mathfrak{o}^{\star}(4)$ symmetries by adding to the results in \cite{BoLuTo2016} three new three-parameter $\mathfrak{o}(2,2)$-real $r$-matrices and one new three-parameter $\mathfrak{o}^{\star}(4)$-real $r$-matrix. Concluding, all nonisomorphic classical $r$-matrices for all real forms of $\mathfrak{o}(4;\mathbb{C})$ are presented in the explicite form what is convenient for the quantizations.

The plan of our Addendum is the following. In Sect.~2 we give a short description of  $D=4$ complex orthogonal Euclidean Lie algebra $\mathfrak{o}(4;\mathbb{C})$ and its real forms: Euclidian $\mathfrak{o}(4)=\mathfrak{o}(3)\oplus\bar{\mathfrak{o}}(3)$, quaternionic $\mathfrak{o}^{\star}(4)=\mathfrak{o}(3)\oplus\bar{\mathfrak{o}}(2,1)$, Kleinian $\mathfrak{o}(2,2)=\mathfrak{o}(2,1)\oplus\bar{\mathfrak{o}}(2,1)$, and Lorentz $\mathfrak{o}(3,1)$ Lie algebras. In Sect.~3 we present complete lists of non-equivalent (nonisomorphic) classical $r$-matrices in terms of the  Cartan--Weyl and Cartesian bases for $\mathfrak{o}(4)$, $\mathfrak{o}^{\star}(4)$, $\mathfrak{o}(2,2)$ and as well in terms of the  Cartan--Weyl and canonical bases for the Lorentz algebra $\mathfrak{o}(3,1)$. In Sect.~4 we present Outlook, where we briefly summarize the obtained results and we mention also some applications of the presented results to the deformations of space-time symmetries and string $\sigma$-models.

\setcounter{equation}{0}
\section{Complex $D=4$ Euclidean algebra and its real forms}
In this section we remind a short  necessary information about structure of $D=4$ complex orthogonal Euclidean Lie algebra $\mathfrak{o}(4;\mathbb{C})$ and its real forms. 

The complex orthogonal Lie algebra $\mathfrak{o}(4;\mathbb{C})$ has the chiral decomposition, i.e. it can be expressed as the following direct sum: 
\begin{eqnarray}\label{rf1}
\mathfrak{o}(4;\mathbb{C})\!\!&=\!\!&\mathfrak{o}(3;\mathbb{C})\oplus\bar{\mathfrak{o}}(3;\mathbb{C}), 
\end{eqnarray}
where $\mathfrak{o}(3;\mathbb{C})$ (respectively $\bar{\mathfrak{o}}(3;\mathbb{C})$) is called left (right) chiral subalgebra. The algebra $\mathfrak{o}(4;\mathbb{C})$ has four real forms where three of them (the compact Euclidean algebra $\mathfrak{o}(4)$, the noncompact quaternionic symmetry $\mathfrak{o}^{\star}(4)$ and the noncompact Kleinian algebra $\mathfrak{o}(2,2)$) preserve the chiral decomposition (\ref{rf1}), i.e. they are expressed as the following direct sums of real $\mathfrak{o}(3;\mathbb{C})$ forms $\mathfrak{o}(3)$ and $\mathfrak{o}(2,1)$: 
\begin{eqnarray}\label{rf2}
\mathfrak{o}(4)\!\!&=\!\!&\mathfrak{o}(3)\oplus\bar{\mathfrak{o}}(3),
\\ \label{rf3}
\mathfrak{o}^{\star}(4)\!\!&=\!\!&\mathfrak{o}(3)\oplus\bar{\mathfrak{o}}(2,1). 
\\ \label{rf4}
\mathfrak{o}(2,2)\!\!&=\!\!&\mathfrak{o}(2,1)\oplus\bar{\mathfrak{o}}(2,1).
\end{eqnarray}
One real form, the Lorentz algebra $\mathfrak{o}(3,1)$, which does not preserve the chiral decomposition, will be considered as well (see (\ref{rf11})). We shall use here two most popular bases of Lie algebra  $\mathfrak{o}(4;\mathbb{C})$ and its real forms: the {\it Cartesian} basis and {\it Cartan--Weyl} one. 

Due to the decompositions (\ref{rf1})--(\ref{rf4}) we shall consider these bases only for one factor $\mathfrak{o}(3;\mathbb{C})$ in (\ref{rf1}) and its real forms $\mathfrak{o}(3)$ and $\mathfrak{o}(2,1)$. The Cartesian basis of $\mathfrak{o}(3;\mathbb{C})$ is given by the generators $I_{i}^{}$ $(i=1,2,3)$ with the defining relations:
\begin{eqnarray}\label{rf5}
\begin{array}{rcl}
[I_{i},\,I_{j}]\!\!&=\!\!&\varepsilon_{ijk}I_{k}.
\end{array}%
\end{eqnarray}
If we consider a Lie algebra over $\mathbb{R}$ with the commutation relations (\ref{rf5}) then we get the compact real form $\mathfrak{o}(3):=\mathfrak{o}(3;\mathbb{R})$ with the anti-Hermitian basis $(i=1,2,3)$:
\begin{eqnarray}\label{rf6}
I^{*}_{i}\!\!&=\!\!&-I_{i}\quad{\rm{for}}\;\mathfrak{o}(3).
\end{eqnarray}
The real form $\mathfrak{o}(2,1)$ is given by the formulas $(i=1,2,3)$:
\begin{eqnarray}\label{rf7}
{I_{i}'}^{\dag}\!\!&=\!\!&(-1)^{i-1}I_{i}'\quad{\rm{for}}\;\mathfrak{o}(2,1).
\end{eqnarray}
where the primed generators satisfy the same relations (\ref{rf5}). For the right chiral Cartesian bases we shall use the notations with bar, i.e. $\bar{I}_i$, $\bar{I}_i'$, $(i=1,2,3)$.

In terms of the Cartesian bases the reality conditions for real forms (\ref{rf2})--(\ref{rf4}) and for the Lorentz algebra $\mathfrak{o}(3,1)$ look as follows $(i=1,2,3)$:
\begin{eqnarray}\label{rf8}
&I^{*}_{i}\;=\;-I_{i}^{},\quad\bar{I}^{*}_{i}\;=\;-\bar{I}_{i}^{}\quad{\rm{for}}\;\mathfrak{o}(4),&
\\[2pt]\label{rf9}
&I_{i}^{*}\,=\,-I_{i},\quad(\bar{I}_{i}')^{\dag}\,=\,(-1)^{i-1}\bar{I}_{i}'\quad{\rm{for}}\;\mathfrak{o}^{\star}(4).&
\\[2pt]\label{rf10}
&(I_{i}')^{\dag}\;=\;(-1)^{i-1}I_{i}',\quad(\bar{I}_{i}')^{\dag}\;=\;(-1)^{i-1}\bar{I}_{i}'\quad{\rm{for}}\;\mathfrak{o}(2,2),&
\\[2pt]\label{rf11}
&I_{i}^{\ddag}\;=\;-\bar{I}_{i}^{},\quad\bar{I}_{i}^{\ddag}\;=\;-I_{i}^{}\quad{\rm{for}}\;\mathfrak{o}(3,1).&
\end{eqnarray}
For the description of quantum deformations, in particular for the classification of classical $r$-matrices of the complex Euclidean algebra $\mathfrak{o}(3;\mathbb{C})$ and its real forms $\mathfrak{o}(3)$ and $\mathfrak{o}(2,1)$, it is convenient to use the {\it Cartan--Weyl} (CW) basis of the isomorphic complex Lie algebra $\mathfrak{sl}(2;\mathbb{C})$ and its real forms $\mathfrak{su}(2)$, $\mathfrak{sl}(1,1)$ and $\mathfrak{sl}(2,\mathbb{R})$. %In the case of $\mathfrak{o}(3)$ the $\mathfrak{su}(2)$ Cartan--Weyl basis can be chosen as follows  
The Cartan--Weyl basis of  $\mathfrak{sl}(2;\mathbb{C})$ can be chosen as follows  
\begin{eqnarray}
\begin{array}{rcl}\label{rf12}
&H\,:=\,\imath I_{3},\quad E_{\pm}\,:=\,\imath I_{1}\mp I_{2},& 
\\[5pt]
&[H,E_{\pm}]\,=\,\pm E_{\pm},\quad[E_{+},E_{-}]\,=\,2H.&
\end{array}%
\end{eqnarray} 
In case of the real form $\mathfrak{o}(3)$ we use the CW basis of $\mathfrak{su}(2)$ which satisfies the relations (\ref{rf12}) with the additional reality condition:
\begin{eqnarray}\label{rf13}
H^{*}\!\!&=\!\!&H,\quad E_{\pm}^{*}\,=\,E_{\mp}^{},
\end{eqnarray}%
where the conjugation ($^*$) is the same as in (\ref{rf6}).

For the real form $\mathfrak{o}(2,1)$ we will use two CW bases of $\mathfrak{sl}(2;\mathbb{C})$ real forms: $\mathfrak{sl}(1,1)$ and $\mathfrak{sl}(2,\mathbb{R})$. Such bases are given by
\begin{eqnarray}
\begin{array}{rcl}\label{rf14}
&H'\,:=\,\imath I_{2}',\quad E_{\pm}'\,:=\,\imath I_{1}'\pm I_{3}',&
\\[3pt]
&[H',E_{\pm}']\,=\,\pm E_{\pm}',\quad[E_{+}',E_{-}']\,=\,2H'
\end{array}{\rm{for}}\;\mathfrak{su}(1,1),&
\\[5pt]
\begin{array}{rcl}\label{rf15}
&H''\,:=\,\imath I_{3}',\quad E_{\pm}''\,:=\,\imath I_{1}'\mp I_{2}', &
\\[3pt]
&[H'',E_{\pm}'']\,=\,\pm E_{\pm}'',\quad[E_{+}'',E_{-}'']\,=\,2H''
\end{array}{\rm{for}}\;\mathfrak{sl}(2,\mathbb{R}).&
\end{eqnarray}
Both bases $\{E_{\pm}',H'\}$ and $\{E_{\pm}'',H''\}$ satisfy the same commutation relations but they have different reality properties, namely
\begin{eqnarray}\label{rf16}
&{H'}^{\dag}\;=\;H',\quad{E_{\pm}'}^{\dag}\;=\;-E_{\mp}'\quad{\rm{for}}\;\mathfrak{su}(1,1),&
\\[2pt]\label{rf17}
&{H''}^{\dag}\;=\;-H'',\quad{E_{\pm}''}^{\dag}\;=\;-E_{\pm}''\quad{\rm{for}}\;\mathfrak{sl}(2;\mathbb{R}),&
\end{eqnarray}
where the conjugation ($^{\dag}$) is the same as in (\ref{rf7})\footnote{It should be noted that in the case of $\mathfrak{su}(1,1)$ the Cartan generator $H$ is compact while for the case $\mathfrak{su}(2,\mathbb{R})$ the generator $H'$ is noncompact.}. 

In the basis (\ref{rf12})--(\ref{rf17}) all possible real forms of $\mathfrak{o}(4;\mathbb{C})$ are described by the following reality conditions:
\begin{eqnarray}\label{rf18}
&H^{*}=H,\quad E_{\pm}^{*}=E_{\mp},\quad\bar{H}^{*}=\bar{H},\quad\bar{E}_{\pm}^{*}=\bar{E}_{\mp}\quad{\rm{for}}\;\mathfrak{o}(4),
\\[5pt]
&\begin{array}{l}\label{rf19}
{H}^{*}=H,\quad E_{\pm}^{*}=E_{\mp},\quad(\bar{H}')^{\dag}=\bar{H}',\quad(\bar{E}_{\pm}')^{\dag}=-\bar{E}_{\mp}',
\\[2pt]
H^{*}=H,\quad E_{\pm}^{*}=E_{\mp},\quad(\bar{H}'')^{\dag}=-\bar{H}'',\quad(\bar{E}_{\pm}'')^{\dag}=-\bar{E}_{\pm}''
\end{array}\;\;{\rm{for}}\;\mathfrak{o}^{\star}(4),
\\[5pt]
&\begin{array}{l}\label{rf20}
{H'}^{\dag}=H',\quad {E_{\pm}'}^{\dag}=-E_{\mp}',\quad(\bar{H}')^{\dag}=\bar{H}',\quad(\bar{E}_{\pm}')^{\dag}=-\bar{E}_{\mp}',
\\[2pt]
{H'}^{\dag}=H',\quad {E_{\pm}'}^{\dag}=-E_{\mp}',\quad(\bar{H}'')^{\dag}=-\bar{H}'',\quad(\bar{E}_{\pm}'')^{\dag}=-\bar{E}_{\pm}'',
\\[2pt]
{H''}^{\dag}=-H'',\quad{E_{\pm}''}^{\dag}=-E_{\pm}'',\quad(\bar{H}'')^{\dag}=-\bar{H}'',\quad(\bar{E}_{\pm}'')^{\dag}=-\bar{E}_{\pm}''
\end{array}{\rm{for}}\;\mathfrak{o}(2,2),
\\[5pt]\label{rf21}
&H^{\ddag}=-\bar{H},\quad E_{\pm}^{\ddag}=-\bar{E}_{\pm},\quad
\bar{H}^{\ddag}=-H,\quad\bar{E}_{\pm}^{\ddag}=-E_{\pm}\quad{\rm{for}}\;\mathfrak{o}(3,1).
\end{eqnarray}

\setcounter{equation}{0}
\section{Classical $r$-matrices of the $\mathfrak{o}(4;\mathbb{C})$ real forms}
In previous paper \cite{BoLuTo2016} we found a total list of nonequivalent (nonisomorphic - unrelated by automorphisms) classical $r$-matrices for the complex $D=4$ Euclidean Lie symmetry $\mathfrak{o}(4;\mathbb{C})$. This result are presented in the form of three three-parameter, one two-parameter and one one-parameter $r$-matrices as follows\footnote{In comparison to \cite{BoLuTo2016} here we remove one three-parameter $\mathfrak{o}(4;\mathbb{C})$ $r$-matrix connected with $r_{2}$ by $\mathfrak{o}(4;\mathbb{C})$ automorphism which permutes the components $\mathfrak{o}(3;\mathbb{C})$ and $\bar{\mathfrak{o}}(3;\mathbb{C})$.}:
\begin{eqnarray}\label{cr1}
%\begin{array}{rcl}
&&r_{1}^{}(\gamma,\bar{\gamma},\eta)=\gamma E_{+}\wedge E_{-}+\bar{\gamma}\bar{E}_{+}\wedge\bar{E}_{-}+\eta H\wedge\bar{H},
\\[1pt]\label{cr2}
&&r_{2}^{}(\gamma,\bar{\chi},\bar{\chi}')=\gamma E_{+}\wedge E_{-}+\bar{\chi}\bar{E}_{+}\wedge\bar{H}+\bar{\chi}'H\wedge\bar{E}_{+},
\\[1pt]\label{cr3}
&&r_{3}^{}(\chi,\bar{\chi},\chi')=\chi E_{+}\wedge H+\bar{\chi}\bar{E}_{+}\wedge\bar{H}+\chi'E_{+}\wedge\bar{E}_{+},
\\[1pt]\label{cr4}
&&r_{4}^{}(\gamma,\chi')=\gamma (E_{+}\wedge E_{-}-\bar{E}_{+}\wedge\bar{E}_{-}-2H\wedge\bar{H})+\chi'E_{+}\wedge\bar{E}_{+},
\\[1pt]\label{cr5}
&&r_{5}^{}(\chi)=\chi(E_{+}+\bar{E}_{+})\wedge(H+\bar{H}).
%\end{array}
\end{eqnarray}
Here all parameters $\gamma$, $\bar{\gamma}$, $\eta$, $\chi$, $\bar{\chi}$, $\chi'$, $\bar{\chi}'$  are arbitrary complex numbers. It should be noted that all three parameters in $r_{1}$ are effective, i.e. there does not exist any $\mathfrak{o}(4;\mathbb{C})$ automorphisms which can reduce the number of parameters. In the case of $r_{2}$ only two parameters are effective because the parameter $\bar{\chi}$ or $\bar{\chi}'$ can be removed by the \textit{rescaling} automorphism $\varphi(\bar{E}_{\pm})=\lambda^{\pm1}\bar{E}_{\pm}$, $\varphi(\bar{H})=\bar{H}$. In $r_{3}$ only one parameters is effective because any two parameters can be removed by the rescaling automorphisms. Analogously, in $r_{4}$ and $r_{5}$ the parameters $\chi'$ and $\chi$ can be removed, i.e. they can be replaced by one.\footnote{Here and further we will not remove the non-effective paramiters because they may be for convenient in quantization procedure.}

In \cite{BoLuTo2016} we employed to (\ref{cr1})--(\ref{cr5}) the reality conditions (\ref{rf18})--(\ref{rf21}) without ones, which contain $\mathfrak{su}(1,1)$ conjugations. In such a way we obtained the classical $r$-matrices for all possible real forms of $\mathfrak{o}(4;\mathbb{C})$: compact Euclidean $\mathfrak{o}(4)$, noncompact quaternionic $\mathfrak{o}^{\star}(4)$, noncompact Kleinian $\mathfrak{o}(2,2)$ and noncompact Lorentz $\mathfrak{o}(3,1)$ Lie symmetries. It follows from \cite{LuTo2017,Zak1994} that the obtained in \cite{BoLuTo2016} sets of the classical $r$-matrices for $\mathfrak{o}(4)$ and $\mathfrak{o}(3,1)$ are complete, whereas for $\mathfrak{o}^{\star}(4)$ and $\mathfrak{o}(2,2)$ these results are only partial. If we apply to (\ref{cr1})--(\ref{cr5}) all conditions (\ref{rf18})--(\ref{rf21}) including ones which contain $\mathfrak{su}(1,1)$ conjugations then we obtain the following results.

I. {\textit{\large Classical $r$-matrices of the real Euclidean algebra $\mathfrak{o}(4)$}}. If we employ the reality conditions (\ref{rf18}) to the results (\ref{cr1})--(\ref{cr5}) we obtain all nonequivalent classical $r$-matrices for the compact Euclidean symmetry $\mathfrak{o}(4)$. These classical $r$-matrices are described by one $*$-anti-Hermitian three-parameter $r$-matrix:
\begin{eqnarray}
\begin{array}{rcl}\label{cr6}
r(\gamma,\bar{\gamma},\eta)\!\!&=\!\!&\gamma E_{+}\wedge E_{-}+\bar{\gamma}\bar{E}_{+}\wedge\bar{E}_{-}+\eta\imath H\wedge\bar{H}
\\[3pt]
\!\!&=\!\!&2\gamma\imath I_{1}\wedge I_{2}+2\bar{\gamma}\imath\bar{I}_{1}\wedge\bar{I}_{2}-\eta\imath I_{3}\wedge\bar{I}_{3}
\end{array}
\end{eqnarray}
with three real parameters $\gamma,\bar{\gamma},\eta$, and $H^{*}=H$, $E_{\pm}^{*}=E_{\mp}$, $\bar{H}^{*}=\bar{H}$, $\bar{E}_{\pm}^{*}=\bar{E}_{\mp}$, $I_{i}^{*}=-I_{i}$, $\bar{I}_{i}^{*}=-\bar{I}_{i}$ ($i=1,2,3$). It should be noted that all parameters in (\ref{cr6}) are effective, i.e. the number of parameters can not be reduced by any $\mathfrak{o}(4)$ automorphism. 

II. {\textit{\large Classical $r$-matrices for the quaternionic algebra $\mathfrak{o}^{\star}(4)$}}. Applying the realty conditions (\ref{rf19}) to the complex formulas (\ref{cr1})--(\ref{cr5}) we obtain all nonisomorphic classical $r$-matrices for the quaternionic algebra $\mathfrak{o}^{\star}(4)$. These classical $r$-matrices are described by three three-parameter anti-Hermitian $r$-matrices: 
\begin{eqnarray}
&\begin{array}{rcl}\label{cr7}
r_{1}^{}(\gamma,\bar{\gamma},\eta)\!\!&=\!\!&\gamma E_{+}\wedge E_{-}+\bar{\gamma}\bar{E}_{+}'\wedge\bar{E}_{-}'+\eta\imath H\wedge\bar{H}',
\\[3pt]
\!\!&=\!\!&2\gamma\imath I_{1}\wedge I_{2}-2\bar{\gamma}\imath\bar{I}_{1}'\wedge\bar{I}_{3}'-\eta\imath I_{3}\wedge\bar{I}_{2}',
\end{array}
\\[4pt]
&\begin{array}{rcl}\label{cr8}
r_{2}^{}(\gamma,\bar{\gamma},\eta)\!\!&=\!\!&\gamma E_{+}\wedge E_{-}+\bar{\gamma}\imath\bar{E}_{+}''\wedge\bar{E}_{-}''+\eta H\wedge\bar{H}'',
\\[3pt]
\!\!&=\!\!&2\gamma\imath I_{1}\wedge I_{2}-2\bar{\gamma}\bar{I}_{1}'\wedge\bar{I}_{2}'-\eta I_{3}\wedge\bar{I}_{3}',
\end{array}
\\[4pt]
&\;\;\begin{array}{l}\label{cr9}
r_{3}^{}(\gamma,\bar{\chi},\bar{\chi}')=\gamma E_{+}\wedge E_{-}+\bar{\chi}\imath\bar{E}_{+}''\wedge\bar{H}''+\bar{\chi}'H\wedge\bar{E}_{+}''
\\[3pt]
\phantom{a}=2\gamma\imath I_{1}\wedge I_{2}-\bar{\chi}(\imath\bar{I}_{1}'-\bar{I}_{2}')\wedge\bar{I}_{3}'+\bar{\chi}'\imath I_{3}\wedge(\imath\bar{I}_{1}'-\bar{I}_{2}),
\end{array}
\end{eqnarray}
where $r_{1}$ supplements the results obtained in \cite{BoLuTo2016}. All parameters $\gamma$, $\bar{\gamma}$, $\eta$, $\bar{\chi}$, $\bar{\chi}'$ are arbitrary real numbers, and $H^{*}=H$, $E_{\pm}^{*}=E_{\mp}$, $(\bar{H}')^{\dag}=\bar{H}'$, $(\bar{E}_{\pm}')^{\dag}=-\bar{E}_{\mp}'$, $(\bar{H}'')^{\dag}=-\bar{H}''$, $(\bar{E}_{\pm}'')^{\dag}=-\bar{E}_{\pm}''$, $I_{i}^{*}=-I_{i}$, $(\bar{I}_{i}')^{\dag}=(-1)^{i-1}\bar{I}_{i}'$ ($i=1,2,3$). Moreover all parameters in $r_{1}$ and $r_{2}$ are effective, and in the case of $r_{3}$ only two parameters are effective because the parameter $\bar{\chi}$ or $\bar{\chi}'$ can be removed by the $\mathfrak{sl}(2,\mathbb{R})$-real \textit{rescaling} automorphism: $\varphi(\bar{E}_{\pm}'')=\lambda^{\pm1}\bar{E}_{\pm}''$, $\varphi(\bar{H}'')=\bar{H}''$, where $\lambda$ is real. 

III. {\textit{\large Classical $r$-matrices for the Kleinian algebra $\mathfrak{o}(2,2)$}}. If we employ the reality conditions (\ref{rf20}) to the complex $r$-matrices (\ref{cr1})--(\ref{cr5}) we obtain all nonisomorphic $\mathfrak{o}(2,2)$-real classical $r$-matrices for the Kleinian symmetry $\mathfrak{o}(2,2)$. These classical $r$-matrices are described by six three-parameter, one two-parameter and one one-parameter nonequivalent $\dag$-anti-Hermitian $r$-matrices:
\begin{eqnarray}
&&\begin{array}{rcl}\label{cr10}
r_{1}^{}(\gamma,\bar{\gamma},\eta)\!\!&=\!\!&\gamma E_{+}'\wedge E_{-}'+\bar{\gamma}\bar{E}_{+}'\wedge\bar{E}_{-}'+\eta\imath H'\wedge\bar{H}'
\\[3pt]
\!\!&=\!\!&-2\gamma\imath I_{1}'\wedge I_{3}'-2\bar{\gamma}\imath\bar{I}_{1}'\wedge\bar{I}_{3}'-\eta\imath I_{2}'\wedge\bar{I}_{2}',
\end{array}
\end{eqnarray}
\begin{eqnarray}
%\\[4pt]
&&\begin{array}{rcl}\label{cr11}
r_{2}^{}(\gamma,\bar{\gamma},\eta)\!\!&=\!\!&\gamma E_{+}'\wedge E_{-}'+\bar{\gamma}\imath\bar{E}_{+}''\wedge\bar{E}_{-}''+\eta H'\wedge\bar{H}''
\\[3pt]
\!\!&=\!\!&-2\gamma\imath I_{1}'\wedge I_{3}'-2\bar{\gamma}\bar{I}_{1}'\wedge\bar{I}_{2}'-\eta I_{2}'\wedge\bar{I}_{3}',
\end{array}
%\end{eqnarray}
%\begin{eqnarray}
\\[4pt]
&&\begin{array}{l}\label{cr12}
r_{3}^{}(\gamma,\bar{\chi},\bar{\chi}')=\gamma E_{+}'\wedge E_{-}'+\bar{\chi}\imath\bar{E}_{+}''\wedge\bar{H}''+\bar{\chi}' H'\wedge\bar{E}_{+}''
\\[3pt]
\phantom{a}=-2\gamma\imath I_{1}'\wedge I_{3}'-\bar{\chi}(\imath\bar{I}_{1}'-\bar{I}_{2}')\wedge\bar{I}_{3}'+\bar{\chi}'\imath I_{2}'\wedge(\imath\bar{I}_{1}'-\bar{I}_{2}'),
\end{array}
%\end{eqnarray}
%\begin{eqnarray}
\\[4pt]
&&\begin{array}{rcl}\label{cr13}
r_{4}^{}(\gamma,\bar{\gamma},\eta)\!\!&=\!\!&\gamma\imath E_{+}''\wedge E_{-}''+\bar{\gamma}\imath\bar{E}_{+}''\wedge\bar{E}_{-}''+\eta\imath H''\wedge\bar{H}''
\\[3pt]
\!\!&=\!\!&-2\gamma I_{1}'\wedge I_{2}'-2\bar{\gamma}\bar{I}_{1}'\wedge\bar{I}_{2}'-\eta\imath I_{3}'\wedge\bar{I}_{3}',
\end{array}
%\end{eqnarray}
%\begin{eqnarray}
\\[4pt]
&&\begin{array}{l}\label{cr14}
r_{5}^{}(\gamma,\bar{\chi},\bar{\chi}')=\gamma\imath E_{+}''\wedge E_{-}''+\bar{\chi}\imath\bar{E}_{+}''\wedge\bar{H}''+\bar{\chi}'\imath H''\wedge\bar{E}_{+}''
\\[3pt]
\phantom{a}=-2\gamma I_{1}'\wedge I_{2}'-\bar{\chi}(\imath\bar{I}_{1}'-\bar{I}_{2}')\wedge\bar{I}_{3}'-\bar{\chi}'I_{3}'\wedge(\imath\bar{I}_{1}'-\bar{I}_{2}'),
\end{array}
\\[4pt]
&&\begin{array}{l}\label{cr15}
r_{6}^{}(\chi,\bar{\chi},\chi')=\chi\imath E_{+}''\wedge H''+\bar{\chi}\imath\bar{E}_{+}''\wedge\bar{H}''+\chi'\imath E_{+}''\wedge\bar{E}_{+}''
\\[3pt]
\phantom{a}=-\gamma(\imath I_{1}'-I_{2}')\wedge I_{3}'-\bar{\chi}(\imath\bar{I}_{1}'-\bar{I}_{2}')\wedge\bar{I}_{3}'+\chi'\imath(\imath I_{1}'-I_{2}')\wedge(\imath\bar{I}_{1}'-\bar{I}_{2}'),
\end{array}
\\[4pt]
&&\begin{array}{l}\label{cr16}
r_{7}^{}(\gamma,\chi')=\gamma\imath (E_{+}''\wedge E_{-}''-\bar{E}_{+}''\wedge\bar{E}_{-}''-2H''\wedge\bar{H}'')+\chi'\imath E_{+}''\wedge\bar{E}_{+}''
\\[3pt]
\phantom{a}=-2\gamma(I_{1}'\wedge I_{2}'-\bar{I}_{1}'\wedge\bar{I}_{2}'-\imath I_{3}'\wedge\bar{I}_{3}')+\chi'\imath(\imath I_{1}'-I_{2}')\wedge(\imath\bar{I}_{1}'-\bar{I}_{2}'),
\end{array}
\\[4pt]
&&\begin{array}{rcl}\label{cr17}
r_{8}^{}(\chi)\!\!&=\!\!&\chi\imath(E_{+}''+\bar{E}_{+}'')\wedge(H''+\bar{H}'')
\\[3pt]
\!\!&=\!\!&-\chi(\imath I_{1}'+\imath\bar{I}_{1}'-I_{2}'-\bar{I}_{2}')\wedge(I_{3}'+\bar{I}_{3}'),
\end{array}
\end{eqnarray}
where $r_{1}$, $r_{2}$ and $r_{3}$ supplement the results obtained in \cite{BoLuTo2016}. All parameters $\gamma$, $\bar{\gamma}$, $\eta$, $\chi$, $\bar{\chi}$, $\chi'$, $\bar{\chi}'$ are arbitrary real numbers, and $(H')^{\dag}=H'$, $(E_{\pm}')^{\dag}=-E_{\mp}'$, $(\bar{H}')^{\dag}=\bar{H}'$, $(\bar{E}_{\pm}')^{\dag}=-\bar{E}_{\mp}'$, $(H'')^{\dag}=-H''$, $(E_{\pm}'')^{\dag}=-E_{\pm}''$, $(\bar{H}'')^{\dag}=-\bar{H}''$, $(\bar{E}_{\pm}'')^{\dag}=-\bar{E}_{\pm}''$, $(I_{i}')^{\dag}=(-1)^{i-1}I_{i}'$, $(\bar{I}_{i}')^{\dag}=(-1)^{i-1}\bar{I}_{i}'$ ($i=1,2,3$). Moreover all parameters in $r_{1}$, $r_{2}$ and $r_{4}$ are effective. In $r_{3}$ and $r_{5}$ only two parameters are effective, because the parameter $\bar{\chi}$ or $\bar{\chi}'$ can be removed by the $\mathfrak{sl}(2,\mathbb{R})$-real \textit{rescaling} automorphism: $\varphi(\bar{E}_{\pm}'')=\lambda^{\pm1}\bar{E}_{\pm}''$, $\varphi(\bar{H}'')=\bar{H}''$, where $\lambda$ is real. In the case $r_{6}$ only one parameter is effective because any two of them  can be removed by the $\mathfrak{sl}(2,\mathbb{R})$-real rescaling automorphisms. Analogously for $r_{7}$ and $r_{8}$ the parameters $\chi'$ and $\chi$ can be removed in the same way.

IV. {\textit{\large Classical $r$-matrices for the Lorentz algebra $\mathfrak{o}(3,1)$}}. Applying the reality conditions (\ref{rf21}) to the complex formulas (\ref{cr1})--(\ref{cr5}) we obtain all nonisomorphic classical $r$-matrices for the Lorentz algebra $\mathfrak{o}(3,1)$. These classical $r$-matrices are described by one three-parameter, two two-parameter and one one-parameter nonequivalent $\ddag$-anti-Hermitian $r$-matrices\footnote{In the calculation of the $r$-matrix (\ref{cr19}) from the formula (\ref{cr3}) we used $\mathfrak{o}(3,1)$-real rescaling automorphism $\varphi(E_{\pm})=\imath\beta^{\pm1}\chi^{\mp1}E_{\pm}$, $\varphi(H)=H$, $\varphi(\bar{E}_{\pm})=-\imath\beta^{\pm1}(\chi^*)^{\mp1}\bar{E}_{\pm}$, $\varphi(\bar{H})=\bar{H}$ where $\beta$ is real and $\chi$ is a complex number.}:  
\begin{eqnarray}
&&\begin{array}{l}\label{cr18}
r_{1}(\alpha,\beta,\eta)=
(\alpha+\imath\beta)E_{+}\wedge E_{-}-(\alpha-\imath\beta)\bar{E}_{+}\wedge\bar{E}_{-}+\eta H\wedge\bar{H}
\\[3pt]
=\displaystyle\frac{\alpha}{2}\imath(e_{+}^{}\wedge e_{-}'+e_{+}'\wedge e_{-}^{})+\frac{\beta}{2}\imath(e_{+}\wedge e_{-}-e_{+}'\wedge e_{-}')-\frac{\eta}{2}\imath h\wedge h',
\end{array}
\\[5pt]
&&\begin{array}{rcl}\label{cr19}
r_{2}(\beta,\chi')\!\!&=\!\!&\beta\imath(E_{+}\wedge\,H+\bar{E}_{+}\wedge\bar{H})+\chi' E_{+}\wedge\bar{E}_{+}
\\[3pt]
\!\!&=\!\!&\displaystyle\frac{\beta}{2}\imath(e_{+}^{}\wedge h-e_{+}'\wedge h')-\frac{\chi'}{2}\imath e_{+}^{}\wedge e_{+}', 
\end{array}
\\[5pt]
&&\begin{array}{l}\label{cr20}
r_{3}(\gamma,\chi')=
\gamma\left(E_{+}\wedge E_{-}-\bar{E}_{+}\wedge\bar{E}_{-}-2H\wedge\bar{H}\right)+\chi'E_{+}\wedge\bar{E}_{+}
\\[3pt]
\phantom{aaaaaaa}=\displaystyle\frac{\gamma}{2}\imath(e_{+}^{}\wedge e_{-}'+e_{+}'\wedge e_{-}^{}+2h\wedge h')-\frac{\chi'}{2}\imath e_{+}^{}\wedge e_{+}'),
\end{array}
\end{eqnarray}
\begin{eqnarray}
%\\[5pt]
&&\begin{array}{l}\label{cr21}
r_{4}(\chi)=\chi\imath(E_{+}+\bar{E}_{+})\wedge(H+\bar{H})=\chi\imath e_{+}\wedge h,
\end{array}
\end{eqnarray}
with the real parameters $\alpha,\beta,\eta,\gamma,\chi,\chi'$, and $H^{\ddag}=-\bar{H}$, $E_{\pm}^{\ddag}=-\bar{E}_{\pm}$, $\bar{H}^{\ddag}=-H$, $\bar{E}_{\pm}^{\ddag}=-E_{\pm}^{}$. Using suitable $\mathfrak{o}(3,1)$-real rescaling automorphisms we can remove the parameter                    $\beta$ in $r_{2}(\beta,\chi')$, both parameters $\gamma$ and $\chi'$ in $r_{3}(\gamma,\chi')$, and the parameter $\chi$ in $r_{4}(\chi)$. In (\ref{cr18})--(\ref{cr21}) the \textit{so-called} canonical basis $\{h,e_{\pm}^{},h',e_{\pm}'\}$ of the Lorentz symmetry $\mathfrak{o}(3,1)$ is related with CW basis $\{H,E_{\pm}^{},\bar{H},\bar{E}_{\pm}\}$ as follows:
\begin{eqnarray}
\begin{array}{rcl}\label{cr22}
h\!\!&=\!\!&H+\bar{H},\qquad\quad e_{\pm}^{}\;=\;E_{\pm}^{}+\bar{E}_{\pm}^{},
\\[3pt]
h'\!\!&=\!\!&-\imath(H-\bar{H}),\quad\; e_{\pm}'\;=\;-\imath(E_{\pm}^{}-\bar{E}_{\pm}^{}).
\end{array}
\end{eqnarray} 
The canonical basis satisfy the following non-vanishing commutation relations:
\begin{eqnarray}
\begin{array}{rcl}\label{cr23}
[h,e_{\pm}^{}]\!\!&=\!\!&\pm e_{\pm}^{},\quad [e_{+}^{},e_{-}^{}]\;=\;2h,
\\[3pt]
[h,e_{\pm}']\!\!&=\!\!&\pm e_{\pm}',\quad [h',e_{\pm}^{}]\;=\;\pm e_{\pm}',\quad [e_{\pm}^{},e_{\mp}']\;=\;\pm2h',
\\[3pt]
[h',e_{\pm}']\!\!&=\!\!&\mp e_{\pm},\quad [e_{+}',e_{-}']\;=\;-2h,
\end{array}
\end{eqnarray} 
and moreover it is anti-Hermitian, i.e.
\begin{eqnarray}\label{cr24}
x^{\ddag}\!\!&=\!\!&-x\quad (x\in\{h,e_{\pm}^{},h',e_{\pm}'\}).
\end{eqnarray} 
The nonisomorphic classical $r$-matrices (\ref{cr18})--(\ref{cr21}) expressed in terms of the generators $\{h,e_{\pm},h',e_{\pm}'\}$ coincide with the Zakrzewski's result \cite{Zak1994} obtained by another method.

\section{Outlook}
We extend the results of the paper \cite{BoLuTo2016} and obtain the complete lists of nonequivalent (i.e. nonisomorhic, unrelated by respective real Lie-algebraic automorphisms) classical $r$-matrices for all real forms of the $D=4$ complex Lie algebra $\mathfrak{o}(4;\mathbb{C})$: Euclidean $\mathfrak{o}(4)$, quaternionic $\mathfrak{o}^{\star}(4)$, Kleinian $\mathfrak{o}(2,2)$ and Lorentz $\mathfrak{o}(3,1)$ Lie algebras. The results are presented in terms of the Cartan--Weyl and Cartesian bases in the case of $\mathfrak{o}(4)$, $\mathfrak{o}^{\star}(4)$, $\mathfrak{o}(2,2)$ as well as in terms of the canonical basis in the case of the Lorentz symmetry $\mathfrak{o}(3,1)$. For the Euclidean real Lie algebra $\mathfrak{o}(4)$ there is only one three-parameter nonequivalent anti-Hermitian $r$-matrix, for quaternionic symmetry $\mathfrak{o}^{\star}(4)$ we obtain three three-parameter anti-Hermitian $r$-matrices, for the Kleinian symmetry $\mathfrak{o}(2,2)$ we find six three-parameter, one two-parameter and one one-parameter nonequivalent anti-Hermitian $r$-matrices, and for the Lorentz symmetry $\mathfrak{o}(3,1)$ there are one three-parameter, two two-parameter and one one-parameter nonequivalent anti-Hermitian $r$-matrices. The completeness of our lists of the classical real $r$-matrices follows from results of the paper \cite{LuTo2017}.

The subsequent problem is to obtain explicit quantizations of the given complex and real bialgebras in the spirit of our papers \cite{LuTo2017,BLT08}. 

In conclusion we recall some potential applications of the obtained results (see \cite{BoLuTo2016}).

\textit{(i)} The $\mathfrak{o}(4)$ $r$-matrix (\ref{cr6}) can be used for the deformations of $S^3$ and $S^2\times S^{2}$ $\sigma$-models and for studying their deformed instanton solutions. We add that the compact spheres $S^{3}$ and $S^2\times S^{2}$ occur also as the internal manifolds in some $D\geq 6$ string theories.

\textit{(ii)} The $\mathfrak{o}^{\star}(4)$ $r$-matrices (\ref{cr7})--(\ref{cr9}) can be used for the construction of YB $\sigma$-models for strings with target spaces $AdS_{2}\times S^{2}$ ($D=4$; see \cite{BBHZZ99,HoTs15}),  $AdS_{2}\times S^{2}\times S^{2}$ ($D=6$; see \cite{AMPPSSTWW15}) and $AdS_{2}\times S^{2}\times T^{6}$ ($D=10$; see \cite{AMPPSSTWW15,Ho14}).

\textit{(iii)} The classical $r$-matrices for the Kleinian algebra $\mathfrak{o}(2,2)$ given by (\ref{cr10})--(\ref{cr17}) describe deformed $D=3$ $AdS$ geometry and can be used for the introduction of YB $\sigma$-models describing the deformations of string models with target spaces $AdS_{3}\times S^{3}$ ($D=6$; see \cite{HoTs15,LRTs14}) and $AdS_{3}\times S^{3}\times S^{3}\times T^{1}$ or $AdS_{3}\times S^{3}\times T^{4}$ ($D=10$; see \cite{AMPPSSTWW15,Ho14,GaGo}). We add that some choices of these $r$-matrices used as deformations of $AdS_3$ were described by Ballesteros et all. \cite{BHMus14}--\cite{BHN15}, obtained by the use of Drinfeld double structures (see \cite{SH02}). 

\textit{(iv)} The $\mathfrak{o}(3,1)$ $r$-matrices (\ref{cr18})--(\ref{cr21}) can be employed for the basic deformations of $D=4$ Lorentz  and $D=3$ de Sitter $(dS)$ symmetry as well as for the deformations of $D=3$ hyperbolic ($H^{3}$) $\sigma$-models.

\subsection*{Acknowledgments}
J.L. would like to thank Stijn van Tongeren for pointing out the insufficiencies of the classification in \cite{BoLuTo2016} of the real $\mathfrak{o}(2,2)$ $r$-matrices. A.B. and J.L. would like to acknowledge the financial support of NCN (Polish National Science Center) grant 2014/13/B/ST2/04043 and by European Project COST, Action MP1405 QSPACE. V.N.T. is supported by RFBR grant No. 16-01-00562-a.

\end{document}